\magnification 1200
\centerline{\bf On ``Observational constraints on Power - Law Cosmologies''}
\vskip 3cm
\centerline {Meetu Sethi, Annu Batra \& Daksh Lohiya}
\centerline {Department of Physics and Astrophysics} 
\centerline {University of Delhi}
\centerline {Delhi 110 007, India}
\centerline {email: meetu@ducos.ernet.in, annu@ducos.ernet.in, dlohiya@ducos.ernet.in}

\vskip 1cm
\centerline{\bf Abstract}
	     "Power Law Cosmologies" are defined by their growth of the
cosmological scale factor  as $ t^\alpha$ regardless of the matter content or
cosmological epoch. Constraints from the current age of the Universe and from
the high redshift supernovae data require ``large'' $\alpha$ ($\approx 1$).
We reinforce this by latest available observations.
Such a large $\alpha$ is also consistent with the right amount of
helium and the lowest observed
metallicity in the universe for a model with the baryon
entropy ratio $\approx 8.1\times 10^{-9}$.

\vfil\eject

     A power law growth, $a(t) = t^\alpha$, for the cosmological scale factor 
$a(t)$, is a generic feature of a class of models that attempt to dynamically 
solve the Cosmological constant
problem [1]. Another example of a power law cosmology is  the 
linear scaling  produced by Allen [6] 
in a  model determined by  an $SU(2)$ cosmological instanton dominated  universe.
As pointed out by Kaplinghat et al [2] (hereinafter referred
to as (I)),  constraints on all such ``power law cosmologies'' from the present
age of the universe and from the high redshift data are consistent with large
$\alpha\approx 1$.
However, (I) considered the primordial light element abundances from early 
universe nucleosynthesis and concluded
that $\alpha$ is forced to lie in a very narrow range
with an upper limit $\approx 0.55$. It was thus concluded in (I) 
that  power-law cosmologies are not viable. 
In this letter, while we reinforce the constraints for a large 
$\alpha\approx 1$ from the more recent data for type Ia supernovae reported
by the supernovae cosmology project [3], we demonstrate that the
nucleosynthesis 
constraints on $\alpha$ arrived at in (I) are seriously in error.  
A large value  $\alpha\approx 1$ 
is consistent with the right amount of helium
observed in the universe in a model with the baryon to entropy ratio
$\approx 8.1\times 10^{-9}$.

     In general for a power law cosmology, the present hubble parameter 
$H_o$ is 
related to the present epoch $t_o$ by $H_o t_o = \alpha$. In what follows 
we shall restrict ourselves to the case where the scale factor 
evolves linearly with time, i.e. $\alpha = 1$. This would include
a Milne cosmology for which $a(t) = t$, as well as a general coasting
cosmology for which $a(t) = kt$ [5]. The Hubble 
parameter is precisely the inverse of the age  $ t_o = (H_o)^{-1}$.
In the standard big bang model [SBB], $ t_o \approx 2/3H_o$.
Thus the age of the universe inferred from a measurement of the Hubble
parameter is 1.5 times the age inferred by the same measurement in
standard matter dominated model. With the best reported value for
$H_o$ standing at $ (H_o) = 100h~ km(sec)^{-1}(Mpc)^{-1}$, with
$ h = 0.65$  [7,8], the age of the universe turns out to
be $ \approx  15 Gyr$. Such an age is comfortably consistent with age
estimates for old clusters.
(I) has put constraints on the value of $\alpha$ using Perlmutter et al [4]
data on SNIa at $ z = 0.83$. The quoted value of 
the figure of merit for SNIa  favours a large  $\alpha\geq 1$.

    For $\alpha = 1$ 
the apparent magnitude $m(z)$, the absolute magnitude $M$ 
and the redshift $z$ of an object 
are related by the exact expression:
$$
  m(z) = 5*log({z^2\over 2} + z) + M - 5*log H_o + 25 
$$
Figure `A' sums up  the  Supernova Cosmology project data 
for supernovae with redshifts
between 0.18 and 0.83 together with the set from the
Calan / Tololo [9] supernovae, at redshifts below 0.1. Figure `B' projects
the data points with the above $m(z)$ curve.
As noted by [3]
the curve for $(\Omega_\Lambda = \Omega_M = 0$ is ``practically identical to 
 $\bf{best fit}$
plot for an unconstrained cosmology''. This reinforces (I) as far as the 
concordance of an $\alpha = 1$ power law cosmology with age and the 
$m - z$ relations are concerned.

	As regards nucleosynthesis, with the expansion scale factor
evolving linearly with time,
the temperature scales as $ aT = tT = $ constant as long as we are in an
era where the photon entropy is not changing much. [The small entropy
change at the time of e+,  e- annihilation does  not alter the following 
argument as well as the results
substantially]. The hubble expansion rate at a given temperature is much 
smaller than its corresponding value at the same temperature in 
standard cosmology. Taking the  present  age as  the inverse  of the
hubble parameter and the present  effective cosmic microwave background (CMB) 
``temperature'' as 2.7 K,
it is easily seen that the universe would be some 50 years old at
temperatures $\approx10^9K$. Such a universe would take some 5000 years to
cool to $10^7K$ !! With the neutron decay rate around 888 seconds at low 
enough temperatures, it would seem that all  neutrons would have decayed
by the time nucleosynthesis may  be  expected to commence  at around
$10^9K$. This is precisely the argument (I) have used to label
nucleosynthesis as spelling ``$disaster$'' for such cosmologies -
and thus ruling them out. 
However, if we consider weak interaction rates of neutrons
and protons, it is easily seen that the  inverse (proton's) beta decay
remains  effective and is not  frozen until temperatures even slightly
less  than $10^9K$. The weak interactions  of the  leptons too remain  in
equilibrium until temperatures even lower: $10^8K$ [10]. 
This has interesting consequences. Firstly, the equality of 
photon and neutrino temperature ($T_\nu = T$) is 
ensured even after the electron - positron annihilation. With temperature
measured in units $T_9 = 10^9K$, this leads to  
an exact expression for the p going to
n rate as $\approx exp[-15/T_9]$ times the n going to p rate.
Figure `C' exhibits the p $\longrightarrow$ n rate in comparison to 
the hubble parameter near $T_9 \approx 1$.
It is clear that by 
inverse beta  decay a proton's conversion  into a neutron is
not decoupled at temperatures as low as $10^9K$. The n/p 
ratio is expected to follow its equilibrium value 
irrespective of the neutron decay rate as long as both n going to p, 
and p going 
to n rates are large in comparison to the expansion rate of 
the universe and the rate of
nucleon leak into the nucleosynthesis channel. 
Although the n/p ratio is small at temperatures $T_9\approx 1$, 
every time any neutron branches off into the nucleosynthesis channel,
the n/p ratio will be replenished by the inverse beta decay of the 
proton.
Simple chemical 
kinetics shows that if we remove one of the reactants or the products
of a reaction in equilibrium at a rate slower than the relaxation period
of the equilibrium buffer, reactions proceed in an equilibrium restoring
direction. As long as we keep precipitating a product at a small enough rate, 
reversible reactions that maintain a solution in equilibrium would restore
the buffer to an equilibrium configuration. This is just what is referred to 
as ``the law of mass action'' in chemistry.

	What actually happens is that, depending on the baryon
entropy ratio, helium starts precipitating out at temperatures around  
$7\times 10^9K$. The rate of precipitation of helium is exhibited
in Figure `D' whence it is clear that the amount of nucleon precipitation into 
helium synthesis channel is negligible in comparison to the neutron formation
and destruction due to inverse and forward beta decay respectively. This is
sufficient to maintain n/p to its equilibrium value. Even in 
(SBB), at such temperatures much higher than the so called deuterium
``bottleneck'' temperature, there is a tiny amount of helium always
forming. However, the universe keeps to such  temperatures in SBB for less than
100's of seconds only and so the amount formed before the``bottleneck'' 
temperature is negligible. In the  case at hand, the universe is at such 
temperatures for some 100 years and the tiny amounts of helium steadily 
builds up. This is conclusively demonstrated by resorting 
to  a numerical  integration of  Boltzmann  equations incorporating
the entire network of reactions. We have
done the required modifications in the standard nucleosynthesis
code outlined by Kawano, to suit the linear expansion of the scale factor. 
Our code integrates 227 
reactions between 64 nuclei and 
takes care of the slow change in baryon entropy ratio
during e+, e- annihilation. 
Runs for different values of baryon to entropy
ratio ($\eta$), and with the currently favoured value of 65 km /sec /Mpc
for the hubble parameter,
yield the result that an
$\eta \approx 8.1\times 10^{-9}$ gives just the 
right amount of Helium [$23.8\%$] 
as observed in the Universe [10].

	As shown in [10], nucleosynthesis in a power law 
cosmology yields a metallicity quite close to the lowest 
observed metallicities. The only problem that one has to contend with is 
the significantly low yields of deuterium in such a cosmology. However,
as pointed out in [10], the amount of Helium produced is quite 
sensitive to $\eta$ in such models. In an inhomogeneous universe, therefore,
one can have the helium to hydrogen ratio to have a large variation. 
Deuterium can be produced by a spallation process much later in the history
of the universe [11]. If one considers spallation of a helium 
deficient cloud onto a helium rich cloud, it is easy to produce deuterium
as demonstrated by Epstein - but without overproduction of Lithium.

	We conclude that nucleosynthesis does not rule out a power law
cosmology as claimed in (I). As a matter of fact it may well turn out to
do better - at least as far as metallicity is concerned. 

	Copies of our numerical code can be downloaded by anonymous ftp
into iucaa.ernet.in. The executable file /in.coming/dlohiya/a.out must be
run on an architecture supporting quadruple precession.

\vskip 1cm
{\bf Figure captions}

Fig. A:  Hubble diagram, the magnitude residual and the uncertainty - normalized
residual plots taken from the supernova cosmology project. {\it ``The curve for
$(\Omega_\Lambda, \Omega_M) = (0,0)$ is practically identical to the best fit 
unconstrained cosmology''}[3].

Fig. B:  The Hubble diagram with the data points (taken from [3])
for linear coasting cosmology.

Fig. C:   The inverse beta decay rate p$\longrightarrow$n and hubble 
	expansion rate as a function of temperature in units of 
	$10^9$ K. Inverse beta decay decouples only at $T_9\approx$ 1.08K.

Fig. D:   Comparison of Helium precipitation and neutron production 
	rates as a function of temperature. Helium production rate, which is 
identically equal to the nucleon precipitation rate out of the n - p 
equilibrium buffer at these temperatures is some 1000 times smaller than 
$n\longleftrightarrow p$ conversion rates by beta decay.

\vfil\eject
{\bf References}
\item{1.} S. Weinberg, Rev. Mod. Phys. 61 (1989); A. D. Dolgov in the 
{\it The Very Early Universe}, eds. G. Gibbons, S. Siklos, S. W. Hawking, 
C. U. Press, 1982; L.H. Ford, Phys Rev D35 (1987) 2339; 
A.D. Dolgov, Phys. Rev. D55 (1997), 5881
\item{2.} Kaplinghat, G. Steigman, I. Tkachev. and T.P. Walker 
astro-ph/9805114, Phys. Rev D59, 043514-[1-5],1999 
\item{3.} S.Perlmutter, et al astro-ph/9812133
\item{4.} S.Perlmutter, et al Nature 391 (1998) 51. 
\item{5.} E.W.Kolb APJ,344, 543, 1989  
\item{6.} R.E.Allen astro-ph/9902042 
\item{7.} D.Branch, Ann.Rev.of Astronomy and Astrophysics 36, 17 (1998),
astro-ph/9801065
\item{8.} W.L. Freedman, J.R. Mould, R.C. Kennicutt,and B.F. Madore, astro-ph
/9801080
\item{9.} M.Hamuy et al, AJ, 1995, 109, 1.
\item{10.}D.Lohiya, A. Batra, S. Mahajan, A. Mukherjee, nucl-th/
          9902022, ``Nucleosynthesis in a Simmering Universe''.
	   	
\item{11.} Epstein, R.I., Lattimere, J.M., and Schramm, D.N.
	  1976, Nature, {\bf 263}, 198

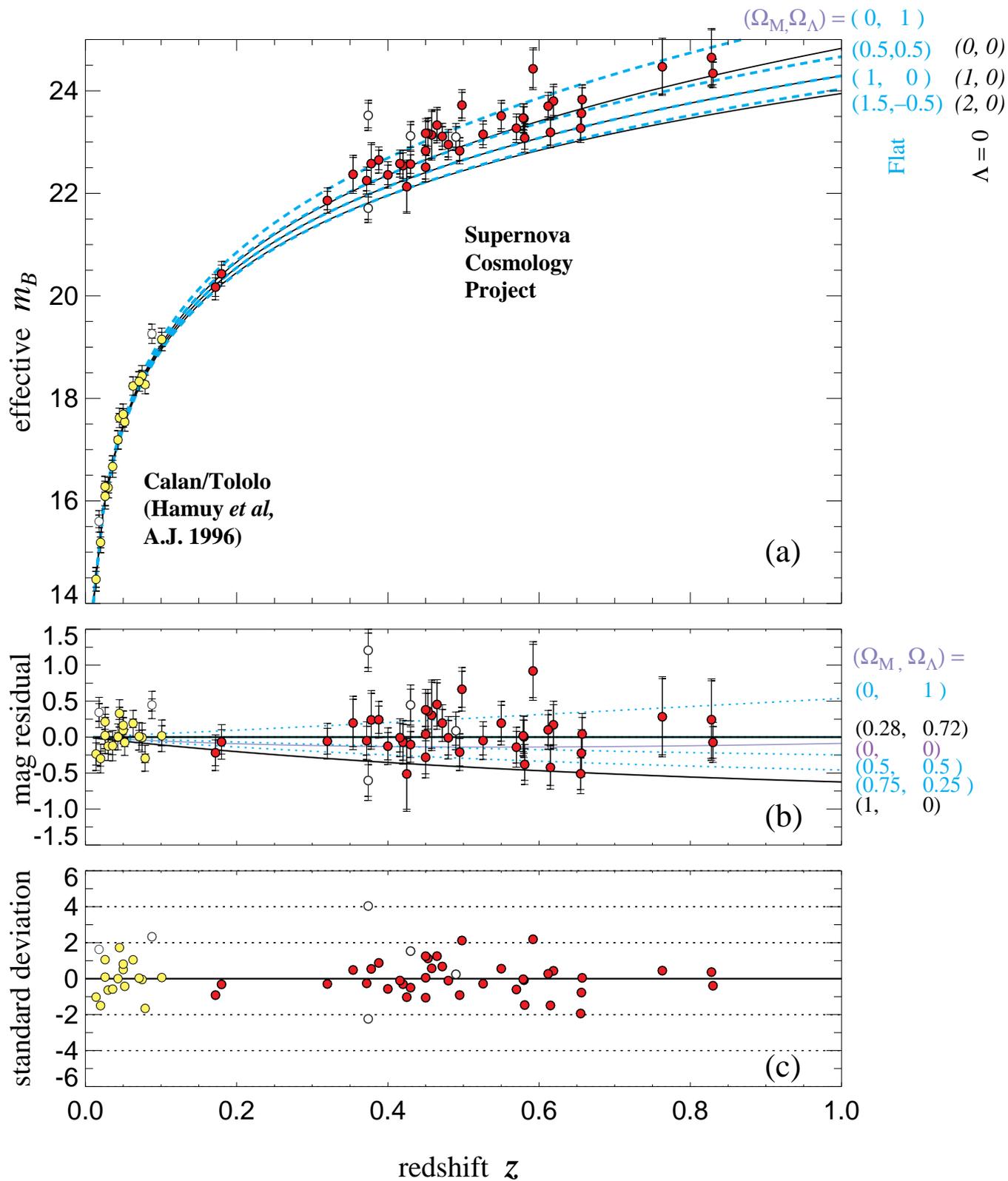
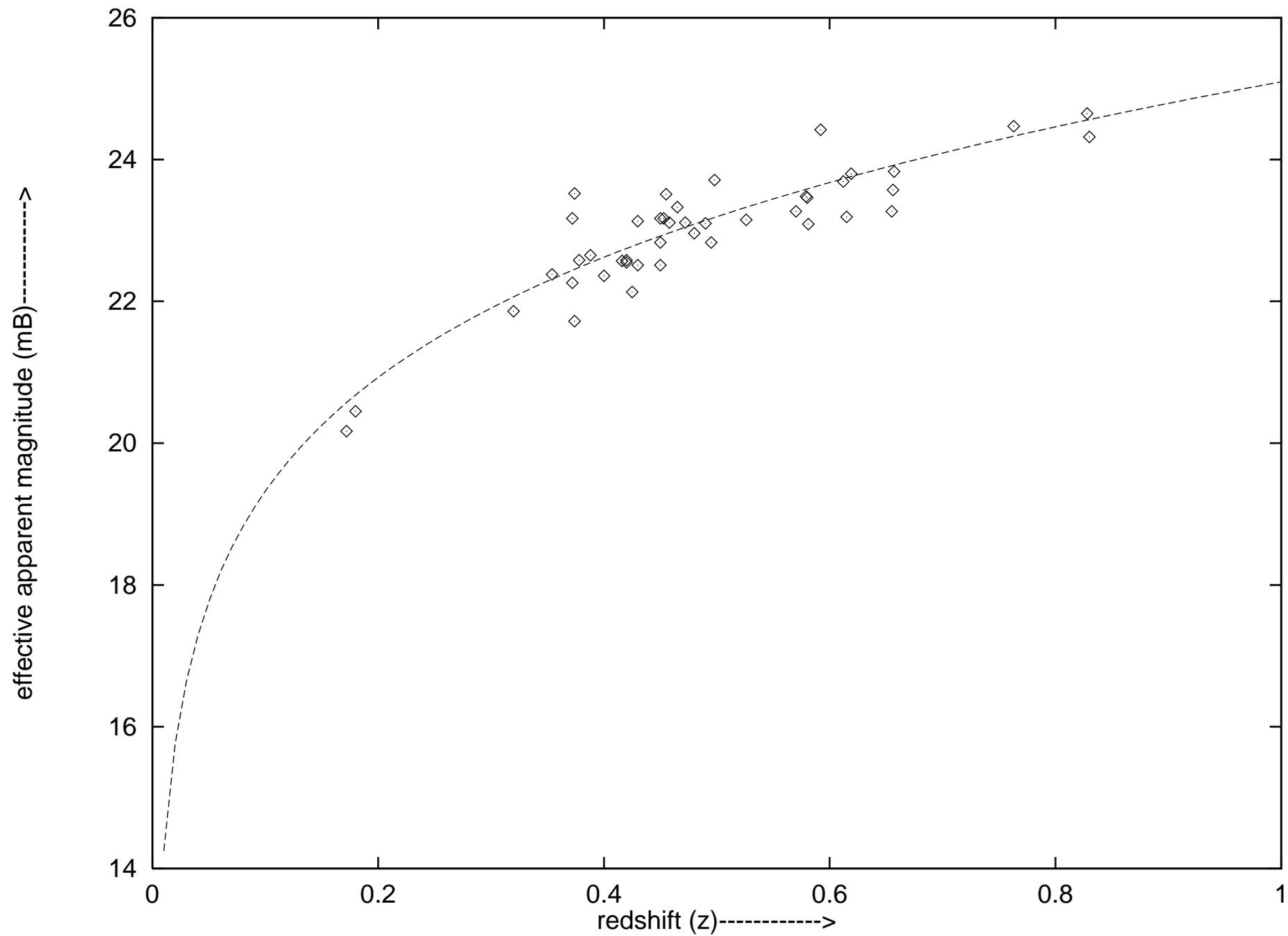
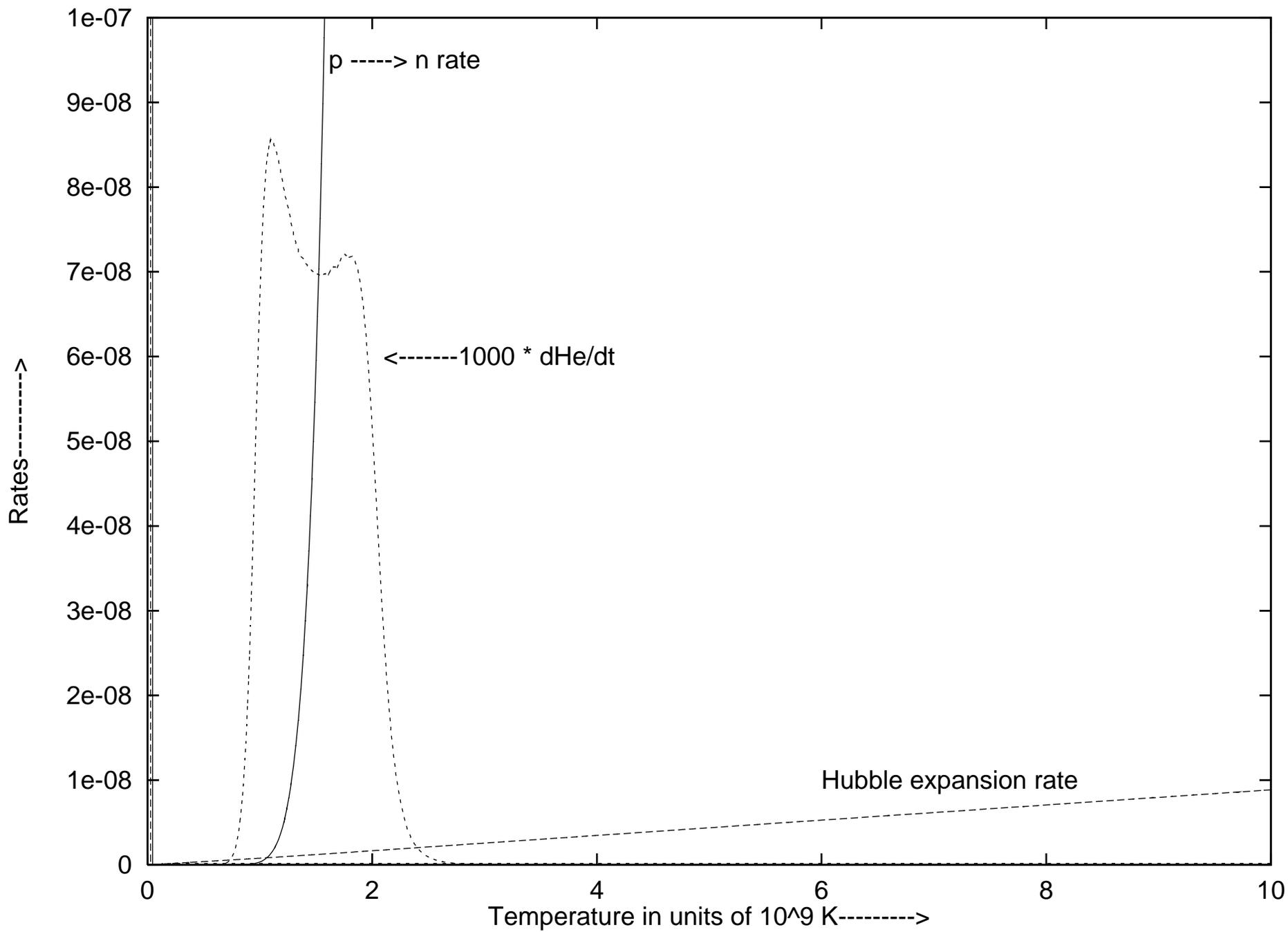
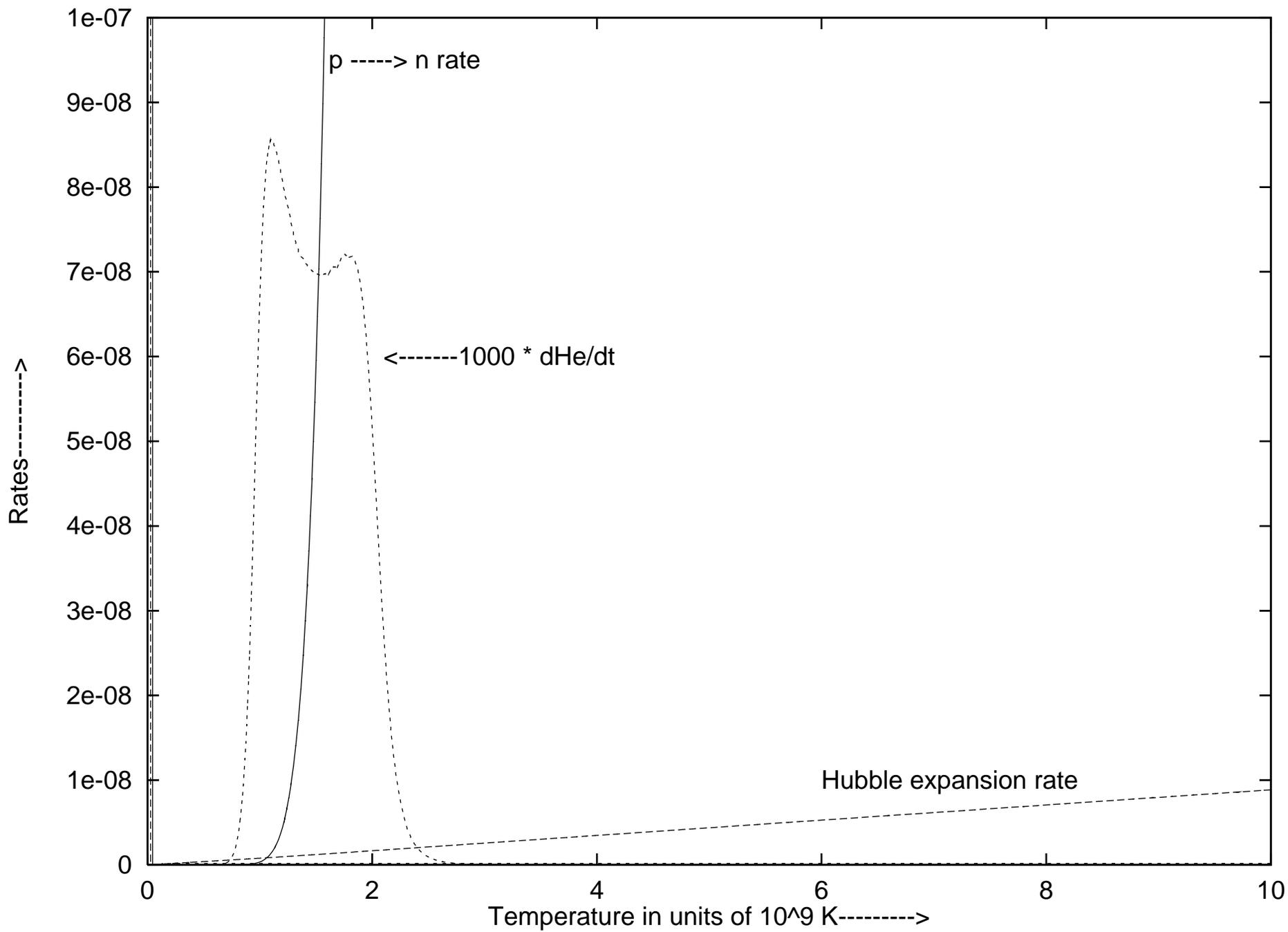
\bye